\documentclass[runningheads]{llncs}
\usepackage{graphicx}
\usepackage{braket}
\usepackage{url}
\usepackage{cite}
\usepackage{amsmath}
\usepackage{color}
\usepackage{lmodern}
\usepackage{siunitx}
\usepackage{booktabs}
\usepackage{etoolbox}
\usepackage{algorithm}
\usepackage{algpseudocode}
\usepackage{diagbox}
\newrobustcmd{\B}{\bfseries}

\begin{document}
\title{QWalkVec: Node Embedding by Quantum Walk}
\author{Rei Sato\inst{}\orcidID{0000-0002-7878-7304} \and
Shuichiro Haruta\inst{}\orcidID{0000-0002-0695-9963} \and
Kazuhiro Saito\inst{}\orcidID{0000-0003-4723-3357} \and
Mori Kurokawa\inst{}\orcidID{0000-0003-4544-0643}} 

%
%\authorrunning{R. Sato et al.}
\institute{AI Division, KDDI Research, Inc.,\\ Fujimino, Ohara 2--1--15, Saitama, 356-8502, Japan
\email{\{ei-satou,sh-haruta,ku-saitou,mo-kurokawa\}@kddi.com}}
\maketitle              % typeset the header of the contribution
\begin{abstract}
In this paper, we propose QWalkVec, a quantum walk-based node embedding method. A quantum walk is a quantum version of a random walk that demonstrates a faster propagation than a random walk on a graph. We focus on the fact that the effect of the depth-first search process is dominant when a quantum walk with a superposition state is applied to graphs. Simply using a quantum walk with its superposition state leads to insufficient performance since balancing the depth-first and breadth-first search processes is essential in node classification tasks. To overcome this disadvantage, we formulate novel coin operators that determine the movement of a quantum walker to its neighboring nodes. They enable QWalkVec to integrate the depth-first search and breadth-first search processes by prioritizing node sampling. We evaluate the effectiveness of QWalkVec in node classification tasks conducted on four small-sized real datasets. As a result, we demonstrate that the performance of QWalkVec is superior to that of the existing methods on several datasets.  Our code will be available at \url{https://github.com/ReiSato18/QWalkVec}.
%138
\end{abstract}
\section{Introduction}

A graph consists of nodes that have labels and attribute information and a set of edges that connect the nodes. For example, the Worldwide Web~\cite{broder2000graph} is a graph in which web pages are nodes, and the hyperlinks between pages are edges. In fields such as social networking services, a node classification problem, where graph structures and node features are used to predict user attributes~\cite{sharma2023deepwalk}, is important. Node classification requires the extraction of features from graph nodes. Node embedding methods have been studied using machine learning to represent the features of each node in a graph as a vector with a fixed length~\cite{perozzi2014deepwalk}.

DeepWalk~\cite{perozzi2014deepwalk} and node2vec~\cite{grover2016node2vec} are famous node embedding techniques that use random walks to sample the node sequences of graphs. DeepWalk consists of two parts: a random walk and Skip-gram~\cite{mikolov2013efficient}. The random walk samples nodes by randomly moving from each node to neighboring nodes, creating a node sequence with the same length as that of the walk. Skip-gram is a neural network that maximizes the co-occurrence probability of the sampled nodes and their surrounding nodes in a given sequence~\cite{mikolov2013distributed}. Node2vec uses biased random walks, which efficiently search the neighborhoods of a given node through tunable parameters used for a breadth-first search (BFS) and a depth-first search (DFS). The BFS prioritizes sampling the neighboring nodes connected to a node, while the DFS aims to sample nodes as far away as possible from the current node. Node2vec enhances the effectiveness of the DFS while maintaining the benefits of the BFS. However, when the input graph is large, the walk length required to acquire sufficient sequence information via a random walk will also be large.

A quantum walk is the quantum version of a random walk and demonstrates faster propagation than a random walk on a graph~\cite{10.5555/1070432.1070590,childs2002example,D.Koch2018}.  Therefore, it has attracted attention as an alternative to random walks. A quantum walk describes the arbitrary direction states in terms of the probability amplitudes of quantum superposition. It exists stochastically at multiple positions at a given walk length. The diffusion rate in one-dimensional space is proportional to the walk length $t$ for the variance of the position $x$ of a random walk, i.e., $\sigma_x^2\sim t$. In contrast, the variance of the position of a quantum walk is proportional to the square of the walk length, i.e., $\sigma_x^2\sim t^2$~\cite{ambainis2001one}.  %Indeed, a quantum walk outperforms a random walk in DFS problems~\cite{D.Koch2018}.

Balancing the BFS and DFS when using quantum walks for node feature representation is also essential in node classification tasks. The reason is that quantum walks are typically used as a superposition state.  If we perform a quantum walk in this superposition state, we confirm that nodes with structural equivalence are preferentially sampled; that is, the DFS effect is dominant, as shown in a later section. In other words, this approach weakens the BFS effect, which is important for learning about neighboring nodes. Therefore, it is necessary to develop a flexible quantum walk that can balance BFS and DFS.

In this study, we propose QWalkVec, a quantum walk-based node embedding algorithm that can balance the BFS and DFS. QWalkVec starts with a superposition state where the DFS effects are optimized, and it incorporates parameters for maximizing the BFS effects with each walk length. Based on node2vec, QWalkVec integrates return and in-out parameters into a coin operator that determines the movement of the quantum walker to its neighboring nodes. QWalkVec creates a unique coin operator for each node, carries out the quantum walk, and generates features for each node. We use four datasets of labeled graphs and compare the node embedding accuracies of QWalkVec, DeepWalk, and node2vec. The contributions of this study are as follows.
\begin{itemize}
\item We formulate a new quantum walk with coin operators that incorporate the effects of the DFS and BFS processes. To the best of our knowledge, this is the first research that tries to formulate a quantum walk for node embeddings.
\item Through extensive experiments, we demonstrate that QWalkVec outperforms DeepWalk and node2vec in several graphs.
\end{itemize}
The rest of this paper is organized as follows. In Section~\ref{sec:preliminaries}, we explain the essential preliminaries. Section~\ref{sec:related work} introduces the related works. In Section \ref{sec:proposed method}, we propose the QWalkVec algorithm for node classification. In Section~\ref{sec:experimental design}, we evaluate the performances of QWalkVec by comparing it with DeepWalk and node2vec. Finally, we conclude the paper in Section~\ref{sec:conclusion}.

\section{Preliminaries\label{sec:preliminaries}}
We first define the notations used throughout the paper. We also present the basic concepts of a quantum walk in this section.

\subsection{Notations}
$G(V,E,M)$ is a labeled undirected graph without edge weights. $G$ consists of a set of $N$ nodes $V=\{1,2,..,i,..,N\}$ and a set of edges $E=\{e_{ij}\}$.  $M$ is the total number of labels.  $\mathrm{\Phi}^{\rm c}\in\mathcal{R}^{N\times d}$ are feature representations of node2vec and DeepWalk, and $\mathrm{\Phi}\in\mathcal{R}^{N\times t}$ are feature representations of QWalkVec.  We write quantum states as $\ket{i\rightarrow j}$ to denote a particle at node $i$ moving toward node $j$. $k_i$ is a degree representing the number of links attached to node $i$.  $t$ is the walk length. $v_0$ is the source node.  For every source node $v_0\in V$, we define $\mathrm{\Phi}^{v_0}$ as a feature representation of node $v_0$ generated through a sampling strategy.

\subsection{Quantum walks on graphs\label{subsec:discrete-time quantum walk}}% one dimenstion?
We define a quantum walk on a graph.  The quantum state of a quantum walk with a length $t$ is defined as
\begin{equation}
    \ket{\psi(t)} = \sum_{i=1}^{N}\sum_{j=1}^{k_i}\psi_{ij}(t)\ket{i}\otimes\ket{i\rightarrow j}.
\end{equation}
$\ket{\psi(t)}$ is defined in the Hilbert space $\mathcal{H} \equiv \mathcal{H}_N\otimes \mathcal{H}_k$, where $\ket{i}\in \mathcal{H}_N$ is associated with the positional degree of freedom, and $\ket{i\rightarrow j} \in \mathcal{H}_k$ is associated with the internal degree of freedom.  $\psi_{ij}(t)$ is the probability amplitude of $\ket{i\rightarrow j}$.  The walk length evolution of the quantum state $\ket{\psi(t)}$ is determined by the coin operator $\hat{C}$ and the shift operator $\hat{S}$:
\begin{equation}
    \ket{\psi(t)} = [\hat{S}\hat{C}]^t\ket{\psi(0)}.
\end{equation}
The coin operator must be a node-dependent since each node has a different number of links, i.e., 
\begin{equation}
    \hat{C} = \sum_i^{N} \ket{i}\bra{i}\otimes\hat{C}_i.
\end{equation}
The coin operator $\hat{C}_i$ at node $i$ is employed as 
$\left(\ket{i}\bra{i}\otimes\hat{C}_i\right)\sum_{j=1}^{k_i}\psi_{ij}(t)\ket{i}\otimes\ket{i\rightarrow j}  = 
    \ket{i}\otimes\hat{C}_i
    \begin{pmatrix}
        \psi_{ij_1}(t),
        \psi_{ij_2}(t),
        ..,
        \psi_{ij_k}(t)
    \end{pmatrix}^{T}$.  Here, the symbol $T$ indicates the transposition operation. The matrix size of the coin operator $\hat{C}_i$ is $k_i\times k_i$. Note that the numbering of the neighboring nodes $\{j_1, j_2, \cdots, j_{k_i}\}$ is arbitrary.  The shift operator changes the position of the quantum walker based on the movement information of the nearest nodes. We define the shift operator as 
\begin{equation}
    \hat{S}\ket{i}\ket{i\rightarrow j} = \ket{j}\ket{j \rightarrow i}.
    \label{eq:flip-flop}
\end{equation}
%The choice of the shift operator of Eq.~(\ref{eq:flip-flop}) is effective for the social networks with varying node degrees.
The probability of node $i$ is given by
\begin{equation}
    P_i(t) = \sum_{j=1}^{k_i}|(\bra{i}\otimes\bra{i\rightarrow j})\ket{\psi(t)}|^2 = \sum_{j=1}^{k_i}|\psi_{ij}(t)|^2.
    \label{eq:quantum_probability}
\end{equation}

\section{Related Works\label{sec:related work}}
Unsupervised feature learning approaches using random walks have been extensively studied. In DeepWalk, which combines Skip-gram with random walks, node embeddings are established by representing the input graph as documents~\cite{ahmed2018learning,perozzi2014deepwalk}. DeepWalk samples nodes from the graph and transforms the graph into ordered node sequences. Node2vec~\cite{grover2016node2vec}, an extension of DeepWalk, introduces search parameters for graphs, proposing a superior sampling strategy.

A comprehensive overview of the field of quantum machine learning is presented by J. Biamonte et al.~\cite{biamonte2017quantum}. For quantum walks, some studies have propose new quantum graph algorithms. F. Mauro et al.~\cite{faccin2014community} show that the initial state of a quantum walk affects the accuracy of community detection. K. Mukae et al.~\cite{mukai2020discrete} show that implementing a quantum walk on a graph reveals a more explicit community structure than that of a random walk. Y. Wang et al.~\cite{wang2022continuous} show that quantum walks exhibit better in graph centrality distinguishing capabilities than classic algorithms.

\subsection{Problems\label{subsec:problems}}
%ネットワークの表現学習に対して, 様々なサンプリング戦略が考えられ, その結果, 学習される特徴表現方法も異なる[node2vec].  実際, すべてのネットワークおよびすべての予測タスク全体に機能する明確に優れたサンプリング戦略に議論の余地がある[node2vec].  グラフをランダムウォークや量子ウォークでサンプリングすることはグラフ上の探索問題と考えられる.  グラフ上の空間探索では, BFS (幅優先検索) と DFS (深さ優先検索) が必須の要素である[node2vec].  BFS は, ノード $i$ に接続されている近辺ノードのサンプリングを優先する.  一方, DFSは深さ優先検索であり, ノード $i$ からできるだけ遠く離れたノードに到達することを目的としている.  DFSの文脈において, グラフ上のランダムウォークによる空間探索は, Nサイトグラフに対して$N$時間オーダーかかる.  この問題に対して, node2vec は, 探索空間を調整するパラメーターを導入しBFSの効果を維持しながらDFSの効果を上げている. 
%量子ウォークは, ランダムウォークと比較して優れた探索能力を備えている.  Deepwalkやnode2vecのように, 量子ウォークと機械学習を組み合わせて新しいサンプリング戦略に基づく分類タスクを実現する方法を考えることは自然であり, 今後の量子コンピュータの発展に伴い重要な課題である.  しかしながら、量子ウォークと機械学習を組み合わせて, 具体的なグラフ処理タスクに応用した研究はほとんどない.  したがって, 我々は量子ウォークの優れた探索能力を活用し, ネットワークにおける効率的な表現学習とノード分類を可能にするサンプリング手法を提案する.  
Various sampling strategies have been studied in graph feature representation learning, resulting in different feature representation methods. Indeed, open problems remain regarding the development of a superior sampling strategy that functions effectively across all graphs~\cite{grover2016node2vec}. Quantum walks demonstrate faster propagation than random walks on graphs~\cite{10.5555/1070432.1070590}. As quantum computing advances, it is natural to consider the node embedding process with quantum walks to realize classification tasks based on new sampling strategies. However, the application of quantum walks for node embedding algorithms remains unexplored. Therefore, we propose QWalkVec, which is a quantum walk-based sampling method.

\section{Proposed Method: QWalkVec\label{sec:proposed method}}

\begin{figure}[tb]
    \centering
   \includegraphics[width=110mm]{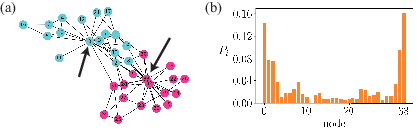}
    \caption{(a) The visualization of Karate dataset. The node indicated by an arrow is a hub node. (b) The quantum probability distribution of each node visited by a quantum walk with the superposition state at $t=100$.}
    \label{fig:karate_dfs_bfs}
\end{figure}
For node representation purposes, homophily and structural equivalence are key concepts~\cite{hoff2002latent}. To capture these characteristics, using both the BFS and DFS strategies is essential~\cite{grover2016node2vec}. The homophily hypothesis~\cite{fortunato2010community} suggests that nodes with similar attributes should belong to the same community and be embedded closely together. BFS involves sampling nodes restricted to the neighbors of a given node, resulting in embeddings that correspond closely to homophily~\cite{grover2016node2vec}.  On the other hand, the structural equivalence hypothesis~\cite{henderson2012rolx} suggests that nodes with similar structural roles in a graph should be closely embedded. Structural equivalence focuses on the structural roles of nodes in graphs.  DFS can sample nodes far from the source node, making it suitable for inferring structural equivalence based on graph roles such as bridges and hubs~\cite{grover2016node2vec}.

To observe how a quantum walk samples nodes, we employ a quantum walk on a Karate~\cite{zachary1977information} as shown in Fig.~\ref{fig:karate_dfs_bfs}.  The initial state of a quantum walk is typically used as a superposition state where the quantum walk starts from all nodes by taking advantage of quantum benefits. When performing a quantum walk in the superposition state, we find that node IDs 0 and 33, which are the hub nodes of the Karate, are almost equally emphasized. Hence, the quantum walk preferentially samples nodes with structural equivalence. On the other hand, the precise information of the neighboring nodes does not appear due to the superposition state. Thus, the BFS effect does not work, and sampling based on a quantum walk tends to have a DFS effect, which extracts structural equivalence.

To overcome this, QWalkVec is proposed. A quantum walk exhibits representation differences depending on the search strategy, such as by changing coin operators~\cite{wong2017coined,mukai2020discrete}. 
Therefore, we propose to leverage coin operators for controlling both BFS and DFS effects.  Fig.~\ref{fig:proposed_method}(a) shows an overview of the QWalkVec.
As shown in this figure, the values on edges defined by $1$, $1/w_p$ and $1/w_q$ are weights of coin operators, which can balance BFS and DFS. 
The return parameter $w_p$ and in-out parameter $w_q$ are created based on each source node, enabling the acquisition of node information around the source node.  When $w_p = w_q = 1$, QWalkVec corresponds to sampling using the conventional quantum walk.  In QWalkVec, we first initiate the initial state from a superposition state, which leverages the DFS effect. The initial state enables all nodes to be searched, resolving the sample size constraint.  We then adjust the priority with which the quantum walker moves to neighboring nodes to characterize the dependencies of nodes.  We repeat sampling for all source nodes and create a feature representation of each node $\mathrm{\Phi}^{v_0}$ shown as Fig.~\ref{fig:proposed_method}(b).  From the summation of each node representation $\mathrm{\Phi}^{v_0}$, we finally create a feature representation $\mathrm{\Phi}$ that captures all node information.  
We describe the detailed procedures in the next section.

\subsection{Algorithm}
\begin{figure*}[tb]
    \centering
    \includegraphics[width=120mm]{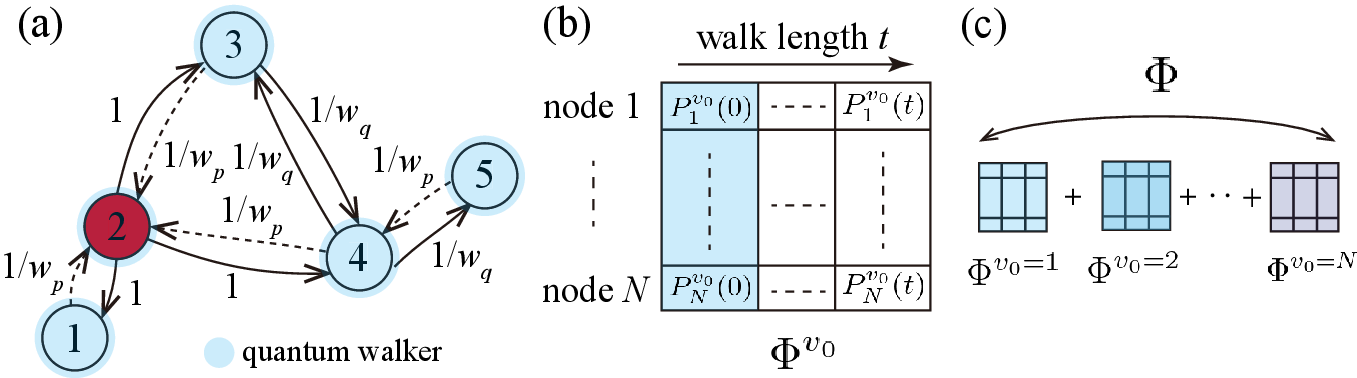}
    \caption{Overview of
QWalkVec. (a) QWalkVec with source node $v_0=2$. (b) $\mathrm{\Phi}^{v_0}$ is the feature representation of node $v_0$. (c) The feature representations $\mathrm{\Phi}$. }
    \label{fig:proposed_method}
\end{figure*}

\begin{algorithm}
\caption{QWalkVec ($G,t,w_p,w_q$)}\label{alg:qw}
\begin{algorithmic}[1]
 \renewcommand{\algorithmicrequire}{\textbf{Input:}}
 \renewcommand{\algorithmicensure}{\textbf{Output:}}
 \Require graph $G(V,E,M)$, 
 \Statex walk length $t$
 \Statex return probability $w_p$
 \Statex in-out probability $w_q$
 \Ensure  matrix of node representations $\mathrm{\mathrm{\Phi}}\in \mathcal{R}^{N\times t}$
 \\ \textrm{Initialization} : set an $N\times t$-size matrix $\mathrm{\mathrm{\Phi}}$
  \For {$v_0 = 1$ to $N$}
  \State set parameter $w_{ij}^{v_0}$ for each node and make $\hat{C}_{v_0}, \ket{\psi(0)}_{v_0}$ (i) 
  \For {$t_i=1$ to $t$}
    \State $\ket{\psi(t)}_{v_0} = [\hat{S}\hat{C}_{v_0}]^{t_i}\ket{\psi(0)}_{v_0}$ (ii)
  \For {$i = 1$ to $N$}
    %\State $P_i^{v_0}(t) =\sum_{j=1}^{k_i}|(\bra{i}\otimes\bra{i\rightarrow j})\ket{\psi(t)}|^2$
    %\State $P_i^{v_0}(t_i)=\sum_{j=1}^{k_i}|(\bra{i}\otimes\bra{i\rightarrow j})\ket{\psi(t)}|^2$
    \State $\mathrm{\Phi}[i,t_i] \leftarrow P_i^{v_0}=\sum_{j=1}^{k_i}|(\bra{i}\otimes\bra{i\rightarrow j})\ket{\psi(t)}|^2$ (ii)
    %\State $\mathrm{\Phi}[i,t_i] \leftarrow P_i^{v_0}(t_i)$
  \EndFor
  \EndFor
  \EndFor
  \\
 \Return $\mathrm{\Phi}$ 
 \end{algorithmic} 
 \end{algorithm}
%Figure~\ref{fig:proposed_method} shows an overview of the QWalkVec approach for node embedding. 
Algorithm~\ref{alg:qw} is a specific implementation. The algorithm is executed as follows: (i) We set the source node $v_0$ and determine coin operators and the superposition state with parameters. (ii) We employ quantum walk, and we obtain the quantum probability $P_i(t)$ and store it in a $N\times t$ matrix. We repeat (i) and (ii) for all $v_0$. For process (i), we define the parameters of the coin operators as
%w_q, w_p逆だわ。重ね合わせ状態からターゲットノードに向けて、w_qじゃないとおかしい。ランダムウォークと逆
\begin{equation}  
w_{ij}^{v_0} = \left\{
\begin{array}{lll}
1 & \text{for} \ket{i}\otimes\ket{i \rightarrow j},& \text{ if } l(v_0, i) = 0 \\
1/w_p & \text{for } \ket{i}\otimes\ket{i \rightarrow j},& \text{ if } l(v_0, i) > l(v_0, j) \\
1/w_q & \text{for } \ket{i}\otimes\ket{i \rightarrow j}, &\text{ if } l(v_0, i) = l(v_0, j)\\
1/w_q & \text{for } \ket{i}\otimes\ket{i \rightarrow j}, &\text{ otherwise}
\end{array}
\right.
\end{equation}
where $l(a,b)$ is the shortest distance between nodes $a$ and $b$. We set $w_p$ as the return probability and $w_q$ as the in-out probability based on node2vec~\cite{grover2016node2vec}. We use the superposition state as the initial state for each source node $v_0$:
\begin{equation}
    \ket{\psi(0)}_{v_0} = \frac{1}{\sqrt{N}}\sum_{i=1}^{N}\frac{1}{\sqrt{\sum_{j=1}^{k_i} w_{ij}^{v_0}}}\sum_{j=1}^{k_i}\sqrt{w_{ij}^{v_0}}\ket{i}\otimes\ket{i\rightarrow j}.
\end{equation}
$w_{ij}^{v_0}$ is a movement weight between nodes $i$ and $j$ for $v_0$. In process (ii), the quantum state at $t$ is defined by 
\begin{equation}
    \ket{\psi(t)}_{v_0} = [\hat{S}\hat{C}_{v_0}]^t\ket{\psi(0)}_{v_0}.
\end{equation}
The coin operator is given by $\hat{C}_{v_0} = \sum_{i}^N\ket{i}\bra{i}\otimes\hat{C}_i^{v_0}$, where $\hat{C}_i^{v_0} = 2\ket{s_i^{v_0}}\bra{s_i^{v_0}} - \hat{I}$.  $\hat{I}$ is identity operator, and $\ket{s_i^{v_0}}$ is given by
\begin{equation}
    \ket{s_i^{v_0}} = \frac{1}{\sqrt{\sum_{j=1}^{k_i} w_{ij}^{v_0}}}\sum_{j=1}^{k_i}\sqrt{w_{ij}^{v_0}}\ket{i\rightarrow j}.
\end{equation}
The shift operator $\hat{S}$ is given by $\hat{S}\ket{i}\ket{i\rightarrow j} = \ket{j}\ket{j \rightarrow i}$. We implement the quantum walk based on the coin operator, which is reconstructed for each $v_0$.

Finally, we define $\mathrm{\Phi}$, a feature representation. Each feature representation $\mathrm{\Phi}^{v_0}$ obtains node information based on each $v_0$ . We assume that the summation of these representations generates an effective composite representation. That is represented as
\begin{equation}
    \mathrm{\Phi}_{i,t} = \sum_{v_0=1}^N \mathrm{\Phi}_{i,t}^{v_0}.
    \label{eq:feature representation}
\end{equation}
The element $\mathrm{\Phi}_{i,t}^{v_0}$ is given by Eq.~(\ref{eq:quantum_probability}) by replacing $\ket{\psi(0)}$ with $\ket{\psi(0)}_{v_0}$. Figs.~\ref{fig:proposed_method}(b) and (c) show the representations of $\mathrm{\Phi}$, $\mathrm{\Phi}^{v_0}$ and Eq.~(\ref{eq:feature representation}). In this study, we investigate the effect of feature addition by combining $N$ normalized feature matrices, ($ \sum_{i=1}^{N} \mathrm{\Phi}_{i,t}^{v_0} = 1$), to form a new feature matrix $\mathrm{\Phi}$ ($ \sum_{i=1}^N \mathrm{\Phi}_{i,t} = N$).

%The feature representation is consist with $\sum_{i=1}^N \mathrm{\Phi}_{i,t} = N$.  Figure~\ref{fig:proposed_method}(b) and (c) show the image of $\mathrm{\Phi}_{i,t}$ and $\mathrm{\Phi}_{i,t}^{v_0}$, respectively.
%Eq.(9)の特徴量を使用する理由は、各ターゲットノードごとに量子ウォークをしたときに、各ウォーク長ごとにノードの重要度を計算するためである。

\section{Evaluations \label{sec:experimental design}}
%\begin{figure*}[tb]
%    \centering
%    \includegraphics[width=100mm]{fig/experimental_setup.eps}
%    \caption{Schematic of our experimental setup. We use three algorithms:(a) DeepWalk (b) node2vec (c) QWalkVec.  ML is a machine learning.}
%    \label{fig:experimental_setup}
%\end{figure*}
%This section overviews the datasets and methods employed for node classification tasks. We first set a labeled graph and then sample nodes from the given graphs to create a node sequence with the length of the random walk or quantum walk. For DeepWalk and node2vec, each sampled node sequence is translated into $d$-dimensional continual values by word2vec. The node sequence sampled by QWalkVec is described by probability, which is a continual value, so we use the probability as a feature learning mechanism without using word2vec. We compare the micro $F_1$ and macro $F_1$ values among the three algorithms.
We evaluate the performance of QWalkVec under the node classification task by numerical simulation, and analyze the parameter sensitivity for the walk length.  Embeddings obtained by each method are fed into a classifier, and it predicts the node labels.  As baselines, DeepWalk~\cite{perozzi2014deepwalk} and node2vec~\cite{grover2016node2vec} are used.

\subsection{Experimental Settings and Dataset}
We follow the experimental procedure of DeepWalk~\cite{perozzi2014deepwalk}.  Specifically, we randomly sample a training set of size $T_R$ from the labeled nodes and use the rest of the nodes as testing data. We repeat this process $20$ times and evaluate the average micro $F_1$ and macro $F_1$ scores as performance scores. We use a one-vs.-rest logistic regression model implemented by LibLinear~\cite{fan2008liblinear} for the node classification tasks of all methods. We present the results obtained by DeepWalk and node2vec with $\gamma=80, w=10$, and $d=128$. We perform a grid search over $p,q,w_p,w_q\in\{0.25,0.50,1,2,4\}$~\cite{grover2016node2vec}.
We use a total of four small-sized real graphs. Karate~\cite{zachary1977information}, Webkb~\cite{nr-aaai15}, Internet Industry Partnerships (IIPs)~\cite{nr-aaai15},  and DD199~\cite{nr-aaai15}.  The statistics of each dataset are shown in the caption of Table~\ref{tab:result}.  
\begin{table}[tb]
    \centering
    \caption{Node classification results in Karate (34 nodes, 78 edges and 2 labels), Webkb (265 nodes and 479 edges and 5 labels), IIPs (219 nodes, 630 edges and 3 labels) and DD199 (841 nodes and 1,902 edges and 20 labels).  We set $(p,q)=(1.0,2.0)$, $(1.0,2.0)$, $(0.25,4.0)$ and $(1.0,0.5)$, and set $(w_p,w_q)=(0.25,1.0)$, $(0.5,4.0)$, $(0.25,4.0)$ and $(1.0,1.0)$ for Karate, Webkb, IIPs and DD199, respectively. Numbers in \textbf{bold} represent the highest performance.}
    \scalebox{0.75}{
    \begin{tabular}{c|c|c|r|r|r|r|r|r|r}
        dataset & & Algorithm \textbackslash $T_R$ & 20\% & 30\% & 40\% & 50\% & 60\% & 70\% & 80\% \\\hline
        Karate & micro $F_1$ (\%)& DeepWalk &&&&87$\pm$23&91$\pm$15&94$\pm$30&96$\pm$33\\
                        && node2vec &&&& \B98$\pm$26&\B98$\pm$28&98$\pm$30&99$\pm$37\\
                     &&QWalkVec&&&&\B 98$\pm$16&\B 98$\pm$18&\B99$\pm$21&\B100$\pm$22\\%\hline
         & macro $F_1$ (\%)& DeepWalk &&&&93$\pm$25&93$\pm$23&93$\pm$32&96$\pm$36\\
                        && node2vec &&&& \B98$\pm$31&\B98$\pm$32&98$\pm$34&99$\pm$38\\
                    && QWalkVec&&&&\B 98$\pm$20&\B 98$\pm$20&\B99$\pm$23&\B100$\pm$22\\\hline
         Webkb&micro $F_1$ (\%)& DeepWalk & 46$\pm$4&48$\pm$3&49$\pm$2&50$\pm$3&51$\pm$2&53$\pm$3&52$\pm$8\\
         &&node2vec & 48$\pm$9&48$\pm$9&49$\pm$9&51$\pm$8&51$\pm$8&53$\pm$8&52$\pm$9\\
                     &&QWalkVec&\B 52$\pm$8 &\B 53$\pm$8&\B 54$\pm$6& \B 55$\pm$6&\B 56$\pm$7&\B 57$\pm$8&\B58$\pm$8\\
         &macro $F_1$ (\%)& DeepWalk&\B29$\pm$2&30$\pm$ 3&31$\pm$5&32$\pm$4&33$\pm$5&35$\pm$3&36$\pm$9\\
         &&node2vec & \B29$\pm$5&29$\pm$5&31$\pm$5&31$\pm$6&32$\pm$7&33$\pm$8&34$\pm$10\\
                    &&QWalkVec&\B 29$\pm$5&\B32$\pm$5&\B33$\pm$5&\B34$\pm$5&\B36$\pm$6&\B37$\pm$7&\B39$\pm$10\\\hline
         IIPs&micro $F_1$ (\%)& DeepWalk &\B65$\pm$3&\B66$\pm$1&\B66$\pm$2&\B67$\pm$4&\B68$\pm$3&69$\pm$3&70$\pm$5\\
         && node2vec &\B65$\pm$6&65$\pm$5&\B66$\pm$5&\B67$\pm$5&\B68$\pm$6&\B70$\pm$8&\B72$\pm$10\\
                     %& QuantumWalk&60$\pm$2&61$\pm$2&62$\pm$2&62$\pm$3&63$\pm$3&64$\pm$4&67$\pm$6\\
                     && QWalkVec&63$\pm$12&64$\pm$9&65$\pm$6&\B67$\pm$5&\B68$\pm$6&\B70$\pm$7&71$\pm$9\\
         &macro $F_1$ (\%)& DeepWalk&\B52$\pm$0&\B52$\pm$0&53$\pm$0&55$\pm$0&\B56$\pm$1&55$\pm$1&56$\pm$1\\
         && node2vec &49$\pm$7&\B52$\pm$6&\B54$\pm$6&\B56$\pm$6&\B56$\pm$7&\B58$\pm$9&\B58$\pm$11\\
                    %& QuantumWalk&42$\pm$5&45$\pm$4&48$\pm$4&48$\pm$4&50$\pm$5&52$\pm$6&54$\pm$6\\
                    &&QWalkVec&40$\pm$7&44$\pm$7&46$\pm$8&49$\pm$8&50$\pm$9&52$\pm$10&57$\pm$11\\\hline
        DD199&micro $F_1$ (\%)& DeepWalk &\B11.6$\pm$3.2&11.7$\pm$2.7&\B12.1$\pm$2.0&\B12.1$\pm$1.4&12.0$\pm$1.9&12.2$\pm$2.3&12.4$\pm$2.8 \\
         &&node2vec&11.4$\pm$3.1&\B12.0$\pm$2.7&12.0$\pm$2.0&\B12.1$\pm$1.6&\B12.5$\pm$1.7&\B12.7$\pm$1.9&\B12.8$\pm$2.8 \\
        && QWalkVec&11.0$\pm$2.6&11.2$\pm$1.8&12.0$\pm$1.4&\B12.1$\pm$1.5&12.1$\pm$1.8&\B12.7$\pm$2.3&12.2$\pm$2.6\\
                     %& QuantumWalk($0.5,0.7$)&10$\pm$1&\B11$\pm$0&11$\pm$1&11$\pm$1&11$\pm$0&\B12$\pm$1&\B12$\pm$1\\\hline
         &macro $F_1$ (\%)& DeepWalk&5.3$\pm$1.2&\B5.4$\pm$0.9&\B5.6$\pm$0.8&5.6$\pm$0.6&5.8$\pm$0.9&5.9$\pm$0.9&6.1$\pm$1.4\\
         &&node2vec&\B5.6$\pm$0.9&\B5.4$\pm$1.2&5.4$\pm$1.1&5.5$\pm$1.1&5.4$\pm$1.4&5.6$\pm$1.6&5.9$\pm$1.9 \\
        && QWalkVec&5.0$\pm$1.0&5.2$\pm$1.0&\B5.6$\pm$1.0&\B5.8$\pm$1.2&\B6.1$\pm$1.5&\B6.2$\pm$1.9&\B6.2$\pm$2.0\\\hline
    \end{tabular}
    }
    \label{tab:result}
\end{table}

\subsection{Overall Results}

\subsubsection{node classification}
Table~\ref{tab:result} shows the experimental results in Karate, Webkb, IIPs and DD199.  For the result of Karate, we set the training ratio $T_R$ from $50\%$ to $80\%$, which is a different situation from other datasets.  We set the training ratio $T_R$ from $20\%$ to $80\%$ for the rest of the datasets.  This is because the number of training nodes becomes too small due to the graph size.  We run for $t=400$ to determine the best approach.  QWalkVec consistently performs better than DeepWalk and node2vec.

For the result of WebKb, we run for $t=400$ to determine the best performance among different walk lengths. For the micro $F_1$ and macro $F_1$ scores, QWalkVec consistently performs better than DeepWalk and node2vec.

For the result of IIPs, we run for $t=400$ to determine the best performance. For the micro $F_1$ score, DeepWalk, node2vec and QWalkVec have almost the same performance, but node2vec consistently performs slightly better than DeepWalk and QWalkVec. For the macro $F_1$ score, node2vec consistently performs better than DeepWalk and QWalkVec.

For the result of DD199.  We run for $t=100$ to determine the best performance.
For the micro $F_1$ score, node2vec consistently performs better than DeepWalk and QWalkVec for $T_R\ge50\%$. However, for $T_R=50\%,70\%$, QWalkVec has the same performance as that of node2vec. For the macro $F_1$ score, DeepWalk and node2vec perform better than QWalkVec at $T_R=20\%,30\%$, but QWalkVec consistently performs better than DeepWalk and node2vec at $T_R\ge40\%$.

\subsubsection{Parameter sensitivity}
%karate, webkb, iip, dd199のTr=50のパラメータ値に対するステップ数の画像を合計8枚(micro, macro)貼る.
\begin{figure*}[tb]
    \centering
    \includegraphics[width=120mm]{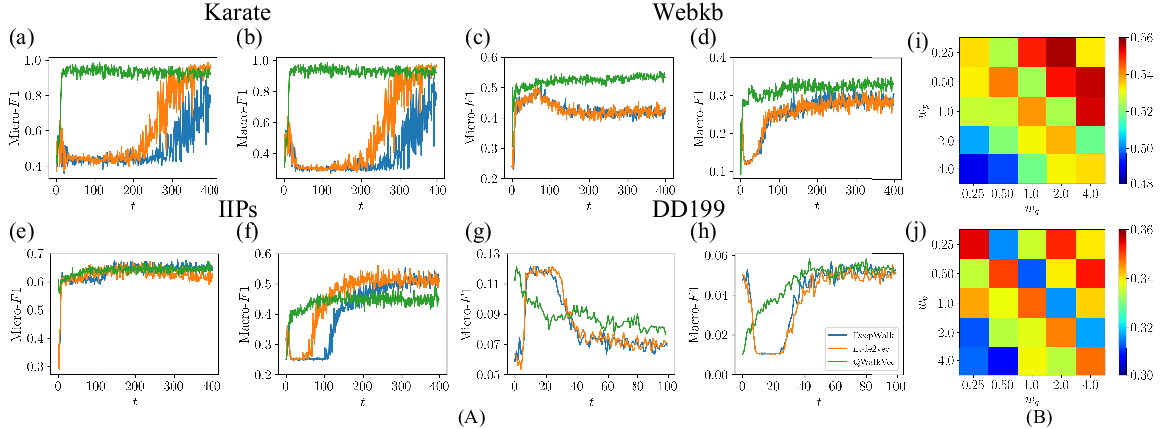}
    \caption{Parameter sensitivity. (A) Walk length dependence of the $F_1$ score with $T_R=50\%$. (B) $F_1$ scores of the $w_p$ and $w_q$ parameters for Webkb with $T_R=50\%$.  We fix the parameter $\gamma=1$ for DeepWalk and node2vec.}
    \label{fig:parameter_sensitivity}
\end{figure*}

QWalkVec involves parameters $t$, $w_p$ and $w_q$. We first examine how the value of $t$ affects the resulting performance. Fig.~\ref{fig:parameter_sensitivity} shows the parameter sensitivity of QWalkVec. We observe that QWalkVec can reach the best performance faster than the DeepWalk and node2vec algorithms since quantum walks demonstrate faster propagation than random walks on graphs. However, the relation between accuracy and the walk length is not correlated for some graphs. For the macro $F_1$ score of IIPs, QWalkVec reaches the best performance faster than DeepWalk and node2vec, but its accuracy is lower than those of these algorithms. The dynamics of the $F_1$ accuracy under different walk lengths exhibit the same pattern for IIPs and DD199 among the three algorithms. For IIPs, we observe that increasing the walk length increases the $F_1$ performance. On the other hand, for the DD199, we observe that increasing the walk length decreases the micro $F_1$ score, while the macro $F_1$ score increases over the walk length $t$. The dynamics of DeepWalk and node2vec also yield the same results for the quantum walk at $t>40$. The walk length dependence of $F_1$ concerns the graph structure. We also examine how different parameter choices affect the performance of the quantum walk on the Webkb dataset using a $T_R=50$ split between the labeled and unlabeled data, as shown in Fig.~\ref{fig:parameter_sensitivity}(B). We measure the macro $F_1$ score as a function of parameters $w_p$ and $w_q$. The performance of QWalkVec tends to improve as the return parameter $w_p$ decreases and the in-out parameter $w_q$ increases, i.e., $w_p<w_q$. While a low $w_q$ encourages DFS, it is balanced by a high $w_p$, which ensures that the quantum walk does not go too far from the starting node. This tendency of the parameter results produced by QWalkVec is similar to the parameter values of node2vec for the used dataset.  

%Discussion
%Karate、Webkbはハブノードをもち、IIPはエッジの偏りがすくない密の高いグラフ、DD199は孤立したノードやグラフが多々あり、スパースなグラフである。結果から、QWalkVecはハブノードがあるグラフに対してDeepWalkやnode2vecよりも精度が高い。macro-F1が低かった、IIPは特徴的なノードが存在しないグラフで
%IIP、DD199 : 一様なグラフに近い：ランダムウォークが小さなクラスにおいても、量子ウォークよりもバランス良く機能している。macro F1スコアはすべてのクラスのパフォーマンスを均等に評価するため、小さなクラスや少数のエッジが関連するパターンが重要あり、量子ウォークは一様なグラフでは細かい情報をとるのが苦手であることが示唆される。KarateやWebkbはハブノードが存在するため、量子ウォークは、特定のノードやパス間の相互作用を利用して効率的に情報を伝達するため、ハブや密度の高いノードが存在するグラフでより優れたパフォーマンスを示した。
\subsubsection{Discussions}
In uniform graphs (IIPs), random walks are more effective for measuring the macro $F_1$ score, which assesses the accuracy achieved for small classes and edges, due to their ability to capture the details of neighboring nodes and connections. On the other hand, in graphs with hubs (Karate and Webkb), the quantum walk, which is more influenced by hubs, attains better results through detailed sampling than those of classic methods. The reason for this is that unlike DeepWalk, which is constrained by the embedding dimensions and window size during the word2vec transformation process, quantum walks directly and probabilistically represent nodes without constraints. In graphs with many missing edges (DD199), there is no significant difference between the proposed method and the existing methods, as the abundance of isolated nodes affects the resulting learning accuracy. The micro $F_1$ score of QWalkVec tends to decrease with the walk length due to the retention of initial state information. However, for the macro $F_1$ score, there is no difference between the proposed approach and the existing methods due to the equivalent sampling process performed on isolated nodes. Consequently, QWalkVec performs superior to DeepWalk and node2vec in graphs with hub nodes.

The time complexity of DeepWalk and node2vec is approximately $\mathcal{O}(N \gamma t_c) + m$, where $m = \mathcal{O}(\log{N})$ represents the time complexity of Skip-gram~\cite{perozzi2014deepwalk}. On the other hand, QWalkVec has a time complexity of $\mathcal{O}(N \cdot t_Q \cdot shots)$, where $shots$ denotes the number of measurements of quantum states.  The actual value of $shots$ is not clear at this time. QWalkVec tends to exhibit a shorter optimal walk length than DeepWalk and node2vec, as illustrated in Fig.~\ref{fig:parameter_sensitivity}, i.e., $t_Q < t_c$.  However, we note that the question of whether QWalkVec will outperform DeepWalk or node2vec with actual quantum computers for sampling time still remains an open problem.

\section{Conclusion \label{sec:conclusion}}
In this study, we propose QWalkVec, a quantum walk-based node embedding algorithm. QWalkVec integrates depth-first search and breadth-first search processes by prioritizing node sampling based on the novel coin operators produced for the quantum walker. We use four labeled graph datasets and compare the node embedding accuracies among QWalkVec, DeepWalk, and node2vec. QWalkVec achieves superior performance to that of DeepWalk and node2vec in graphs with hub nodes. For our future work, we will investigate the performance of QWalkVec in other graph-related tasks such as community detection.

%\bibliographystyle{splncs04}
\bibliographystyle{unsrt}
\bibliography{mybib}

\end{document}